\documentclass[prd,aps,showpacs,floats,floatfix,superscriptaddress]{revtex4}

\usepackage[dvips]{graphicx}

\usepackage{longtable}
\usepackage{dcolumn,epsfig}

\def\laq{\raise 0.4ex\hbox{$<$}\kern -0.8em\lower 0.62ex\hbox{$\sim$}}
\def\gaq{\raise 0.4ex\hbox{$>$}\kern -0.7em\lower 0.62ex\hbox{$\sim$}}

\newcommand{\beq}{\begin{equation}}
\newcommand{\eeq}{\end{equation}}
\newcommand{\bea}{\begin{eqnarray}}
\newcommand{\eea}{\end{eqnarray}}
\newcommand{\ba}{\begin{array}}
\newcommand{\ea}{\end{array}}

\newcommand{\eqref}[1]{(\ref{#1})}

\newlength{\sizeonefig}
\newlength{\sizetwofig}
\newlength{\sizeonefigb}
\newlength{\sizetwofigb}
\renewcommand{\arraystretch}{1}

\begin{document}

\title{\large\bf Quantum noise in laser-interferometer gravitational-wave detectors
 with a heterodyne readout scheme}

\author{Alessandra Buonanno}

\affiliation{Groupe de Gravitation et Cosmologie (GReCO), 
Institut d'Astrophysique de Paris (CNRS),
98$^{\rm bis}$ Boulevard Arago,
75014 Paris, France}

\affiliation{Theoretical Astrophysics, California
Institute of Technology, Pasadena, CA 91125}

\author{Yanbei Chen}

\affiliation{Theoretical Astrophysics, California
Institute of Technology, Pasadena, CA 91125}

\author{Nergis Mavalvala}

\affiliation{Department of Physics, Massachusetts Institute of Technology, Cambridge, MA 02139}

\begin{abstract}
We analyze and discuss the quantum noise in signal-recycled laser
interferometer gravitational-wave detectors, such as Advanced
LIGO, using a heterodyne readout scheme and taking into account
the optomechanical dynamics. Contrary to homodyne detection, a
heterodyne readout scheme can simultaneously measure more than one
quadrature of the output field, providing an additional way of
optimizing the interferometer sensitivity, but at the price of
additional noise. Our analysis provides the framework needed to
evaluate whether a homodyne or heterodyne readout scheme is more
optimal for second generation interferometers from an
astrophysical point of view. As a more theoretical outcome of our
analysis, we show that as a consequence of the Heisenberg
uncertainty principle the heterodyne scheme cannot convert
conventional interferometers into (broadband) quantum
non-demolition interferometers.
\end{abstract}

\maketitle

\section{Introduction}
\label{sec1}

Long-baseline laser-interferometer gravitational-wave (GW)
detectors have begun operation in the United States
(LIGO~\cite{LIGO}), Europe (VIRGO~\cite{VIRGO} and
GEO\,600~\cite{GEO}) and Japan (TAMA\,300~\cite{TAMA}). Even as
the first detectors begin the search for gravitational radiation,
development of the next generation detectors, such as Advanced
LIGO (or LIGO-II), is underway. With planned improvements in the
seismic noise reduction --- via active vibration
isolation~\cite{AdLIGOsei}, and in the limits set by thermal noise
--- via the improved mechanical quality of the optics and clever
suspension strategies~\cite{AdLIGOsus}, the sensitivity of second
generation detectors is expected to be quantum-noise-limited in
much of the detection band from $10$ to $10^4$\,Hz.

The optical configuration of all current GW detectors includes a
Michelson interferometer. Two 4\,km-long Fabry-Perot cavities are
inserted into the arms of the Michelson interferometer; the
optical field builds up in the cavities and samples the GW-induced
phase shift multiple times. The arm cavities, thus, increase the
phase sensitivity of the detector. The Michelson-based optical
configuration makes it natural to decompose the optical fields and
the mechanical motion of the arm-cavity mirrors into modes that
are either symmetric (i.e.\ equal amplitude) or antisymmetric
(i.e.\ equal in magnitude but opposite in signs) in the two arms,
as explained in detail, for example, in
Refs.~\cite{KLMTV00,BC1,BC2,BC3}. No light leaves the
interferometer from below the beam-splitter (BS) or dark port,
except the lights induced by the antisymmetric motion of the
arm-cavity test-mass mirrors, e.g., due to a passing-by
gravitational wave, or due to vacuum fluctuations that originally
enter the interferometer from the dark port. Since GW
interferometers operate on a dark fringe, the intensity of the
light exiting the antisymmetric port is quadratic in the GW
amplitude, and therefore insensitive to it, to first order. The
standard way to circumvent this is to interfere the signal field
with a relatively strong local oscillator (LO) field, such that
the intensity of the total optical field, detected at the beat
frequency, varies linearly with the GW amplitude. The various
methods of measuring the GW-induced signal at the antisymmetric
port are referred to as {\it readout schemes.}

Previously~\cite{KLMTV00,BC1,BC2,BC3,BC4}, the quantum noise in
Advanced LIGO was calculated assuming a homodyne readout scheme,
in which the LO field oscillates at exactly the same frequency as
the incident laser.  The homodyne readout scheme can pose
significant technical challenges for laser noise. In this paper we
consider heterodyne readout schemes, in which the LO has different
frequencies from the carrier. The heterodyne readout is usually
implemented, as in initial LIGO (or LIGO-I), by using phase
modulated light: the light incident on the interferometer consists
of a carrier and radio frequency (RF) phase modulation (PM)
sidebands~\footnote{~Because all cases of heterodyning we consider
in this work are carried out at radio frequencies, we refer to
this readout as {\it RF modulation-demodulation}.}. Using the
Schnupp asymmetry~\cite{Schnupp}, the PM sidebands are transmitted
to the photodetector as efficiently as possible, while the carrier
still returns to the bright port. The transmitted sidebands then
act as a LO against which the GW signal can beat. Demodulation at
the modulation frequency converts the signal back down into the
baseband. This technique circumvents laser technical noise by
upconverting the signal detection to frequencies where the laser
light is shot-noise-limited (a few MHz). \emph{Here we do not
concern ourselves with technical noise on the laser; we consider
only the fundamental quantum noise on the light}. When the RF
modulation-demodulation readout scheme is implemented, more than
one quadrature of the interferometer output will be available for
measurement, providing an additional tool for the optimization of
the sensitivity, which is not available in homodyne detection.

However, an \emph{additional} quantum noise contribution, as
compared with the homodyne readout scheme, usually appears in this
scheme during the photodetection process --- as was realized by
Gea-Banacloche and Luechs in Ref.~\cite{GBL} where they evaluated
the compatibility of squeezing and modulation-demodulation readout
schemes in simple Michelson interferometers, and also by Schnupp,
using more general considerations~\cite{Schnupp}. This additional
contribution is due to vacuum fluctuations in frequency bands that
are twice the modulation frequency away from the carrier.
Subsequently, the heterodyne scheme was investigated in more
detail by Niebauer et al.~\cite{N88} and Meers and
Strain~\cite{MS91}. These works~\cite{GBL,Schnupp,N88,MS91}
focused exclusively on the detection of the output {\it phase}
quadrature with {\it phase} modulated LO light (at the output
port), which is appropriate for conventional GW interferometers
with low circulating power, and hence negligible back action
noise, but not for the advanced GW interferometers considered
here.

The main purpose of this paper is to further generalize the
results obtained in Refs.~\cite{GBL,Schnupp,N88,MS91}, by
including the possibility of detecting {\it generic} quadratures
with LO light that are {\it mixtures} of phase and amplitude
modulation to the carrier, and applying them to advanced GW
interferometers, such as Advanced LIGO. In particular, we provide
expressions and examples of the quantum noise, taking into account
explicitly both the variable-quadrature optimization and the
additional heterodyne noise. This lays the foundation for
optimization of the detector sensitivity for specific
astrophysical sources and for comparison between heterodyne and
homodyne schemes from an astrophysical point of view. The results
of these investigations will be reported elsewhere ~\cite{BCM2}.

Recently, Somiya~\cite{Somiya} showed independently the
possibility of measuring different quadratures through heterodyne
detection, and investigated the consequences for both conventional
and signal-recycled interferometers. However, the additional
heterodyne noise was not explicitly taken into account --- it was
hoped that, in certain sophisticated heterodyne schemes, the
additional heterodyne noise becomes negligible, while the
variable-quadrature optimization remains possible. However, as we
show in this paper, the additional heterodyne noise is a direct
consequence of the Heisenberg uncertainty principle, and will
always exist as long as more than one quadrature is available for
simultaneous measurement. Moreover, the Heisenberg uncertainty
principle gives rise to a quantum limit to the additional
heterodyne noise, which is frequency independent unless a
frequency-dependent squeezing is implemented. This
frequency-independent quantum limit will seriously constrain the
power of the variable-quadrature optimization of heterodyne
schemes in achieving (broadband) quantum non-demolition (QND)
performances. In fact, for conventional interferometers, all
quantum-limited heterodyne detection can be shown to be equivalent
to a frequency-independent homodyne detection performed on an
otherwise identical conventional interferometer with lower input
laser power.
\begin{figure}
\begin{center}
\epsfig{file=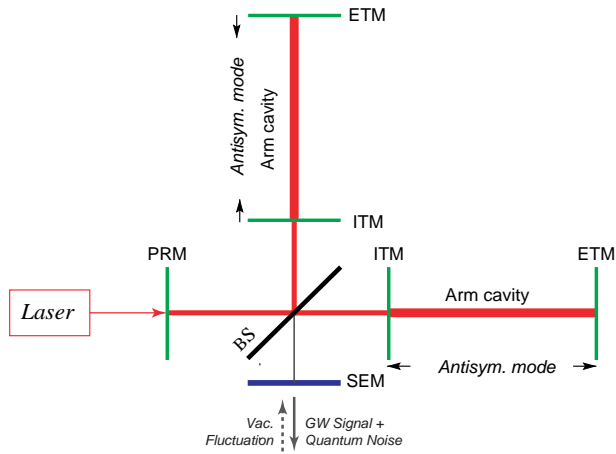,width=\sizeonefig} \caption{We draw a
signal- (and power-) recycled LIGO interferometer. The laser light
enters the interferometer from the left (bright port), through the
power-recycling mirror (PRM), and gets split by a 50/50
beamsplitter into the two identical (in absence of gravitational
waves) arm cavities. Each of the arm cavities is formed by the
input test-mass (ITM) and the end test-mass (ETM) mirrors. A
signal extracting mirror (SEM) is placed at the dark port, forming
a signal extracting (SE) cavity with the ITMs.} \label{Fig1}
\vspace{-0.6cm}
\end{center}
\end{figure}

This paper is organized as follows: in Section~\ref{sec2} we
describe the modulation/demodulation process and derive the
demodulated output of signal-recycled interferometers in terms of
quadrature operators and arbitrary heterodyne field amplitudes ---
taking the simplest sine-wave modulation/demodulation scheme as an
example; in Section~\ref{sec3} we derive the expressions of the
quantum noise spectral density in this scheme, and apply them to
initial and Advanced LIGO interferometers;  in
Section~\ref{sec4.1} we analyze a completely general
modulation/demodulation scheme, derive a quantum limit for
heterodyne measurements and discuss the consequence of this
quantum limit for conventional interferometers. Finally, in
Section~\ref{sec5}, we present our conclusions.

\section{The Radio-Frequency modulation-demodulation scheme in advanced LIGO}
\label{sec2}

\subsection{Overview of Advanced LIGO optical configuration}
\label{sec2.1}

The Michelson interferometer is operated on the dark fringe to
minimize static laser power, and hence the shot noise associated
with this light, at the antisymmetric (dark) port. Since most of
the light returns toward the laser, a partially transmitting
mirror, the power-recycling mirror (PRM) is placed between the
laser and the beam splitter to `recycle' the light back into the
interferometer~\cite{Drever} (see Fig.~\ref{Fig1}). The optical
configuration currently planned to achieve quantum-limited
performance in Advanced LIGO uses the Resonant Sideband Extraction
(RSE) technique~\cite{RSE}, in addition to power-recycling. In
RSE, an additional partially transmitting mirror, the signal
extraction mirror (SEM), is placed between the antisymmetric port
of the beamsplitter and photodetector (see Fig~\ref{Fig1}).

The optical properties (reflectivity, loss) of this signal
extraction mirror and its microscopic position (in fractions of
the wavelength of the laser light, $1.064\,\mu{\rm m}$) can
significantly influence the frequency response of the
interferometer~\cite{SR,RSE}. When the signal extraction cavity
(SEC) --- comprising the SEM and the input test-mass (ITM) mirrors
of the arm cavities --- is exactly resonant or anti-resonant at
the laser frequency, the bandwidth of the entire detector can be
increased or decreased by altering the reflectivity of the SEM.
These two special cases are referred to as resonant sideband
extraction (RSE)~\cite{RSE} and signal recycling (SR)~\cite{SR},
respectively.

As the signal cavity is slightly offset (detuned) from resonance
(RSE) or antiresonance (SR), the frequency at which the peak
optical response of the detector occurs can be shifted to
frequencies where other noise sources are not dominant. Note that,
unlike conventional interferometers and tuned RSE/SR
interferometers, the frequency responses of detuned configurations
are no longer symmetric around the carrier frequency, with only
one resonant peak located either higher or lower than the carrier
frequency. As a consequence, although the interferometer will
respond resonantly to GW's with a certain nonzero frequency, only
one of the two (upper and lower) sidebands the GW generates
symmetrically around the carrier frequency is on resonance. More
generally, the upper and lower GW sidebands contribute
asymmetrically to the total output field, which makes the GW
signal appear simultaneously in both quadratures of the output
field~\cite{BC1,BC2,BC3,BC4}. Detuned configurations are neither
RSE nor SR in the original sense, but roughly speaking, such a
configuration can be classified as either RSE or SR by looking at
whether the bandwidth of the entire interferometer is broader or
narrower than that of the arm cavity. Historically, since SR was
invented earlier than RSE, some literature refers to all
configurations with a signal mirror as ``Signal Recycled''.

Since detuned RSE allows us to control the spectral response of
the interferometer and optimize for specific astrophysical
sources, it has become a strongly favored candidate for Advanced
LIGO~\footnote{~RSE,
  instead of SR, is chosen for Advanced LIGO in order to decrease the
  required input power~\cite{RSE}. However, as far as quantum noise is concerned, the required circulating power inside
the arms will not be influenced by whether SR or RSE is used
~\cite{RSE,BC4}.}. A notable consequence is that with the high
laser power of Advanced LIGO, the optomechanical coupling induced
by detuned RSE/SR significantly modifies the dynamics of the
interferometer, introducing an additional resonance at which the
sensitivity also peaks~\cite{BC1,BC2,BC3,BC4}.

\subsection{Modulation and demodulation processes}
\label{sec2.2}

The RF modulation-demodulation scheme comprises two parts: the
modulation-preparation and demodulation-readout processes. In this
section we consider only the simplest case, sine wave modulation
and demodulation. A more general discussion of
modulation/demodulation schemes can be found in Sec.~\ref{sec4}.

Phase modulated light is incident on the interferometer.  It is
composed of the carrier at the laser frequency $\omega_0 \sim
2\cdot 10^{15}\,{\rm s}^{-1}$, and a pair of phase modulation
sidebands offset from the carrier by several MHz, so $\omega_{\rm
m} \sim 2 \pi \cdot 10^6\,{\rm s}^{-1} \gg$ GW-sideband frequency
$<2 \pi \cdot 10^4 \,{\rm s}^{-1}$.  The detection port is kept as
dark as possible for the carrier, while the PM sidebands at
$\omega_0\pm\omega_{\rm m}$ are coupled out as efficiently as
possible to act as the local oscillator for the GW-induced carrier
field that leaks out. Maximal RF sideband transmission is adjusted
in two ways: (i) by a path difference in the arms of the Michelson
that is arranged to be highly transmissive for the RF component of
the field -- the Schnupp asymmetry; and (ii) by matching the
transmission of the power-recycling and signal-extracting mirrors
so that the effective cavity comprising those two mirrors is
critically coupled.  In addition to the gravitational-wave
readout, the PM sidebands are also useful for controlling the
auxiliary degrees of freedom of the interferometer~\cite{Srse02}.

\begin{figure}
\begin{center}
\epsfig{file=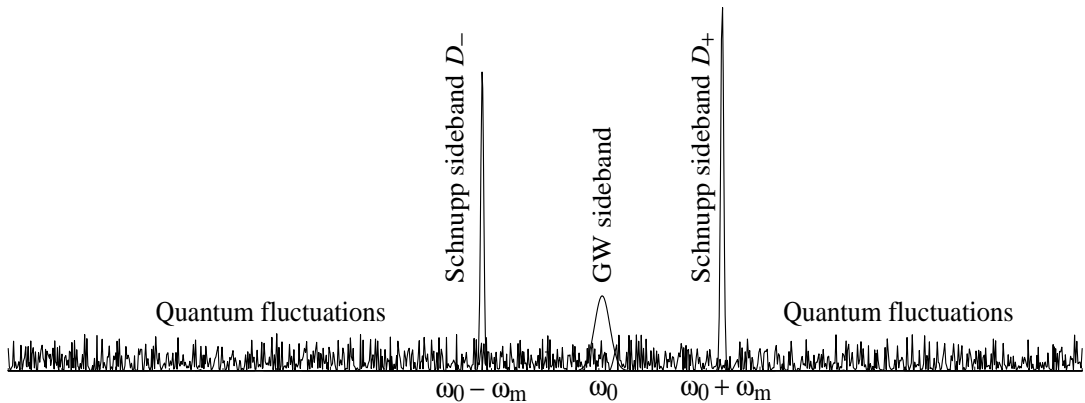,width=0.8\textwidth,height=0.3\textwidth}
\caption{The outgoing field at the interferometer output consists of GW sideband
signals around $\omega_0$, the Schnupp sideband fields at
$\omega_0\pm\omega_{\rm m}$ and quantum fluctuations spread out at
all frequencies.}
\label{Fig2}
\vspace{-0.6cm}
\end{center}
\end{figure}

In Fig.~\ref{Fig2} we show the outgoing optical field at the
interferometer output in the frequency domain, which consists of
the GW sidebands, the Schnupp sidebands and quantum fluctuations
at the output.
The relative amplitudes of the RF sidebands are intentionally
shown to be unequal. This is a case of unbalanced heterodyning,
and is  an intrinsic feature of the detuned RSE interferometer. As
described above, when we move to detuned RSE, the SEC is detuned
from perfect carrier resonance, such that the resonance peak of
the signal cavity coincides with \emph{one} of the GW signal
sidebands, at the expense  of the other one; consequently, the GW
signal appears in both quadratures of the output field. At the
same time, this phase shift in the signal cavity moves both RF
sidebands off perfect resonance as well, which results in poor
output coupling of both RF sidebands. This can be remedied  by
offsetting the RF sideband frequency -- or conversely, the
macroscopic length of the SEC -- to make \emph{one} of the RF
sidebands resonant~\cite{MW}. Hence, detuned RSE leads to
unbalanced heterodyne fields. Although the carrier is phase
modulated before entering the interferometer, the heterodyne
fields at the output port will no longer act as a pure phase
modulation on the carrier.

At the detection port, a standard heterodyne detection procedure is
used to extract the GW signal:
\bea
{\rm Outgoing\;light}\Rightarrow {\rm Photodetection}
\rightarrow
{\rm Mixing\;with}\,\cos(\omega_{\rm m} t + \phi_{\rm D})
\rightarrow
{\rm Low-pass\;filter}
\Rightarrow
{\rm Demodulated\;output} \nonumber
\eea
The photodetection process consists of taking the square of the
optical field shown in Fig.~\ref{Fig2}. This operation mixes the
GW signal (and quantum fluctuations) located at frequency
$\sim\omega_0$ with the RF sideband fields located at frequency
$\omega_0\pm\omega_{\rm m}$. As a consequence the GW signal is
measured in the RF band around $\pm \omega_{\rm m}$. By taking the
product of (or mixing) the photodetection output with the
demodulation function, $\cos(\omega_{\rm m} t +\phi_{\rm D})$, the
GW signal is down-converted back to low frequencies. The result is
then filtered by a low-pass filter, yielding a {\it
frequency-independent} quadrature that does depend on $\phi_{\rm
D}$. However, as we shall see more quantitatively in the following
sections, in addition to the GW signal (and quantum fluctuations)
centered at $\sim \omega_0$, quantum fluctuations at
$\sim\omega_0\pm2\omega_{\rm m}$ also enter the demodulated output
at the antisymmetric port. This gives rise to an additional noise
term that is not present in a homodyne readout scheme.

\subsection{Demodulated output of LIGO interferometers}
\label{sec2.3}

The optical field coming out from the
interferometer [see Fig.~\ref{Fig1}] can be written as a sum of two parts:
\beq
{E}(t)=L(t)+{S}(t)\,.
\label{2.1}
\eeq
The first term,
\beq
L(t) = \left[ D_+ e^{i\,(\omega_0+\omega_{\rm m})\,t}+D_-
e^{i\,(\omega_0-\omega_{\rm m})\,t} \right]  + {\rm h.c.}\,,
\eeq
is the (classical) LO light composed of the Schnupp sideband
fields at frequencies $\omega_0 \pm \omega_{\rm m}$, with
(complex) amplitudes $D_+$ and $D_-$, respectively. The magnitude
and phase of $D_\pm$ depend on the specific optical configuration.
The two quadratures of the LO are generated by either amplitude
modulation (first quadrature) or phase modulation (second
quadrature) of the input light. The second term in Eq.~(\ref{2.1})
,
\bea
{S}(t) &\equiv&
\int_{0}^{+\infty}\frac{d\omega}{2 \pi}
\left[\,
e^{-i \omega t}\,{b}_{\omega}
+{\rm h.c.}
\right]\,,
\nonumber \\
&=&\int_{-\Lambda}^{+\Lambda} \frac{d\Omega}{2\,\pi}\,
\left[\,
e^{-i\,(\omega_0-2\omega_{\rm m}+\Omega)\,t}\, {b}_{\omega_0-2\omega_{\rm m}+\Omega}
+{\rm h.c.}
\right]
\nonumber \\
&+&\int_{-\Lambda}^{+\Lambda} \frac{d\Omega}{2\,\pi}\,
\left[\,
e^{-i\,(\omega_0+\Omega)\,t}\, {b}_{\omega_0+\Omega}
+{\rm h.c.}
\right]
\nonumber \\
&+&\int_{-\Lambda}^{+\Lambda} \frac{d\Omega}{2\,\pi}\,
\left[\,
e^{-i\,(\omega_0+2\omega_{\rm m}+\Omega)\,t}\, {b}_{\omega_0+2\omega_{\rm m}+\Omega}
+{\rm h.c.}
\right]
\nonumber \\
&+&({\rm contributions\;at\;irrelevant\;frequency\;bands})\,, \eea
contains both the (classical) GW signal and the quantum
fluctuations of optical fields near $\omega_0$.
Here $\Lambda \lesssim \omega_{\rm m}$ is the demodulation
bandwidth \footnote{~Note that both in $L(t)$ and ${S}(t)$
we disregard the overall factor $\sqrt{2 \pi\hbar \omega_0/{\cal A} c}$
where ${\cal A}$ is the effective cross sectional area of the laser beam and
$c$ is the speed of light. This factor does not affect the final
expression of the spectral density and for simplicity we neglect it.
See Eq. (2.6) in Ref.~\cite{BC2}.}. [For simplicity and clarity, we only
write out explicitly the terms that will eventually contribute to
the demodulated output.]

The photocurrent from the photodetector is proportional to the square
of the optical field,
\bea
\label{esq}
i(t)\propto {E}^2(t)
&=&L^2(t)+2 L(t)\, {S}(t)+  {S}^2(t)\,, \nonumber \\
&=&
\left[{\rm contributions\;at\;frequencies\;}
0,\,\pm 2\omega_{\rm m},\,\pm (2 \omega_0\pm2\omega_{\rm m}),\;{\rm from\;} L^2(t)
\right] \nonumber \\
&+&
2 D_+ \left[
\int_{-\Lambda}^{+\Lambda} \frac{d\Omega}{2\pi}
e^{i(\omega_{\rm m}-\Omega)t}\,{b}_{\omega_0+\Omega}
+
\int_{-\Lambda}^{+\Lambda} \frac{d\Omega}{2\pi}
e^{-i(\omega_{\rm m}+\Omega)t}\,{b}_{\omega_0+2\omega_{\rm m}+\Omega}
\right] +{\rm h.c.} \nonumber \\
&+&
2 D_- \left[
\int_{-\Lambda}^{+\Lambda} \frac{d\Omega}{2\pi}
e^{i(\omega_{\rm m}-\Omega)t}\,{b}_{\omega_0-2\omega_{\rm m}+\Omega}
+
\int_{-\Lambda}^{+\Lambda} \frac{d\Omega}{2\pi}
e^{-i(\omega_{\rm m}+\Omega)t}\, {b}_{\omega_0+\Omega}
\right] +{\rm h.c.} \nonumber \\
&+&\left[{\rm contributions\;at\;irrelevant\;frequency\;bands}\right]
\nonumber \\
&+& \left[{\rm terms\;quadratic\;in\;} {b},\,{b}^{\dagger},\;{\rm
from\;} S^2(t)\right]\,. \eea
After taking the product of (or mixing) $i(t)$ with
$\cos(\omega_{\rm m} t +\phi_{\rm D})$ and applying a low-pass
filter with cutoff frequency $\Lambda$, we obtain the demodulated
output
\bea
\label{DemoOut}
i(t) \cos (\omega_{\rm m} t +\phi_{\rm D})
\stackrel{\rm Low-pass}{\Longrightarrow}
{O}(\phi_{\rm D};t)
&=&
\int_{-\Lambda}^{+\Lambda} \frac{d\Omega}{2\pi}
\left[
\left( D_+  e^{-i \phi_{\rm D}} +  D_- e^{i \phi_{\rm D}} \right) e^{-i\, \Omega\, t}\,
b_{\omega_0+\Omega} + {\rm h.c.}
\right]
\nonumber \\
&+&
\int_{-\Lambda}^{+\Lambda} \frac{d\Omega}{2\pi}
\left[
D_+ e^{i \phi_{\rm D}} e^{- i\, \Omega\, t}\, {b}_{\omega_0+2 \omega_{\rm m} + \Omega} +{\rm
h.c.} \right] \nonumber \\
&+&
\int_{-\Lambda}^{+\Lambda} \frac{d\Omega}{2\pi}
\left[D_- e^{- i \phi_{\rm D}} e^{- i\, \Omega\, t}\, {b}_{\omega_0 -
2 \omega_{\rm m} + \Omega} +{\rm h.c.}
\right]\,,
\eea
where we assume the local oscillator to be strong enough that the
quadratic terms in Eq.~(\ref{esq}) can be ignored. It is
convenient to recast the demodulated output (\ref{DemoOut})  in
terms of quadrature operators by using the following relation (for
$\Lambda \ll \omega$):
\beq
\label{CAtoQ}
\int_{-\Lambda}^{+\Lambda} \frac{d\Omega}{2\pi}
\left[A e^{-i\,\Omega\,t}\,{b}_{\omega + \Omega} + A^* e^{i\,\Omega\,t}\,
{b}^{\dagger}_{\omega + \Omega}
\right]
=
\int_{0}^{+\Lambda} \frac{d\Omega}{2\pi}
\sqrt{2}\, A_0 \,e^{-i\,\Omega\,t} \, {b}_{\alpha +\frac{\pi}{2}}^{\omega}(\Omega)
+ {\rm h.c.}\,.
\eeq
Here $A=A_0\, e^{i\alpha}$ ($A_0,\,\alpha \in \Re$) is an arbitrary
complex amplitude,  and the quadrature operator
${b}_{\alpha+\frac{\pi}{2}}^{\omega}$ is defined as [see also Refs.~\cite{BC1,BC2}]
\beq
{b}_{\zeta}^{\omega}(\Omega)={b}_1^{\omega}(\Omega) \sin\zeta +
{b}_2^{\omega}(\Omega) \cos\zeta\,,
\eeq
where
\beq
{b}_1^{\omega}(\Omega)=\frac{{b}_{\omega+\Omega}+{b}_{\omega-\Omega}^{\dagger}}{\sqrt{2}}\,,
\quad \quad {b}_2^{\omega}(\Omega)=\frac{{b}_{\omega+\Omega}-
{b}_{\omega-\Omega}^{\dagger}}{\sqrt{2}i}\,.
\eeq
The superscript $\omega$ on the quadrature fields is added to emphasize that
the quadratures are defined with respect to the central frequency $\omega$.
By applying relation (\ref{CAtoQ}) to the demodulated output
(\ref{DemoOut}), we get
\beq
\label{outQ}
{O}(\phi_{\rm D};t)=
\int_{0}^{+\Lambda} \frac{d\Omega}{2\pi}
e^{-i\, \Omega\, t}
\sqrt{2}
\left[
D_0 \, {b}_{\zeta_0}^{\omega_0}(\Omega)
+\left|D_+\right|\,{b}_{\zeta_+}^{\omega_0+2\omega_{\rm m}}(\Omega)
+\left|D_-\right|\, {b}_{\zeta_-}^{\omega_0-2\omega_{\rm m}}(\Omega)
\right] +{\rm h.c.} \,,
\eeq
in which we have defined,
\beq
\label{D_0}
D_0 \equiv \left| D_+\,e^{- i\phi_{\rm D}}+D_-\,e^{i\phi_{\rm D}}\right|\,,
\eeq
and
\bea
\label{zeta_0}
\zeta_0 &\equiv& \frac{\pi}{2}+ \arg\left(D_+\,e^{- i\phi_{\rm D}}+D_-\,e^{i\phi_{\rm
D}}\right)\,,  \\
\zeta_{\pm} &\equiv& \pm \phi_{\rm D} + \frac{\pi}{2} +\arg D_{\pm}\,.
\eea
In the frequency domain, we have,
\beq
\label{outQFD}
{O}(\phi_{\rm D};\Omega)=\sqrt{2}D_0
\left[
{b}_{\zeta_0}^{\omega_0}(\Omega)
+\frac{\left|D_+\right|}{D_0}\,{b}_{\zeta_+}^{\omega_0+2\omega_{\rm m}}(\Omega)
+\frac{\left|D_-\right|}{D_0}\,{b}_{\zeta_-}^{\omega_0-2\omega_{\rm m}}(\Omega)
\right]\,,\quad\quad|\Omega|<\Lambda\,.
\eeq

The first term inside the parenthesis, $b_{\zeta_0}^{\omega_0}$ is
an output quadrature field around the carrier frequency
$\omega_0$, which contains both the GW signal and vacuum
fluctuations in the optical fields near the carrier frequency. In
Refs.~\cite{BC1,BC2,BC3,BC4}, this quadrature field is related to
the input quadrature field at the antisymmetric port via the
input-output relations, from which the spectral density of the
quantum noise can be derived. Measuring this field is the task of
all readout schemes. For example, a homodyne scheme can measure
directly an arbitrary frequency-independent quadrature. For this
reason, we call the quadrature field $b_{\zeta_0}^{\omega_0}$ the
{\it homodyne quadrature} for distinction.  The two additional
terms inside the parenthesis are the additional noise, which come
from vacuum fluctuations around $\omega_0\pm2\omega_m$. The sum of
all three terms is what we measure in the heterodyne scheme, which
we call the {\it heterodyne quadrature}.

\begin{table*}[ht]
\begin{tabular}{ccc}
\hline\hline
Quantity & Symbol and Value  \\
\hline\hline
Laser frequency & $\omega_0=1.8\times10^{15}\,{\rm sec}^{-1}$ \\
GW sideband frequency & $\Omega$  \\
Input test-mass transmissivity & $T$  \\
Arm-cavity length & $L=4\,{\rm km}$\\
Mirror mass & $m$\\
Light power at beamsplitter & $I_0$ \\
SEM amplitude reflectivity and transmissivity & $\rho,\tau$  \\
SEC length & $l\sim10\,{\rm m}$ \\
SEC detuning & $\phi=[\omega_0 l/c]_{\rm mod\,2\pi}$ \\
\hline\hline
\end{tabular}
\caption{Basic quantities of Advanced LIGO interferometers \label{tab:basic}}
\end{table*}

\subsection{Features of the RF modulation-demodulation scheme}
\label{sec2.4}

As can be inferred from Eq.~(\ref{zeta_0}), as long as
$|D_+|\neq|D_-|$, all homodyne quadratures can be measured through
some heterodyne quadrature with the appropriate demodulation phase
$\phi_{\rm D}$. The (single-sided) spectral density $S(\Omega)$
associated with the noise $h_n$ can be computed by the
formula [see Eq.~(22) of Ref. \cite{KLMTV00}]:
\beq
\label{33}
2\pi\,\delta(\Omega - \Omega^\prime)\,S(\Omega)
= \langle {\rm in} | h_n(\Omega)\,h_n^\dagger(\Omega^\prime) +
h_n^\dagger(\Omega^\prime)\,h_n(\Omega)|{\rm in} \rangle \,,
\eeq
and if the input state of the whole interferometer
is the vacuum state ($|{\rm in} \rangle = |0_{{a}} \rangle$),
the following relation holds:
\beq
\label{vac}
 \langle 0_{{a}}| {a}_i(\Omega)\,{a}^\dagger_{j}(\Omega')
+ {a}^\dagger_{j}(\Omega') \,{a}_i(\Omega)
|0_{{a}} \rangle= 2\pi\,\delta(\Omega- \Omega^\prime)\,
\delta_{i j}\,.
\eeq
{}From Eq.~(\ref{outQFD}) we see that the noise spectral density in the heterodyne
quadrature is a sum of that of the homodyne quadrature, $S_b^{\rm
hom}(\phi_{\rm D};\Omega)$, and those of the additional noise terms,
$S_b^{\rm add}(\phi_{\rm D};\Omega)$. Since $S_b^{\rm hom}(\phi_{\rm
D};\Omega)$ and $S_b^{\rm add}(\phi_{\rm D};\Omega)$ come from
different frequency bands, we assume that they are uncorrelated, hence
\beq
\label{Sbhet}
S_b^{\rm het}(\phi_{\rm D};\Omega)=S_{b}^{\rm hom}(\phi_{\rm
D};\Omega) + S_b^{\rm add}(\phi_{\rm D};\Omega)\,.
\eeq
Assuming that the fields associated with the additional heterodyne noise are in
the vacuum state, we get a white (frequency-independent) spectrum
for the additional noise,
\beq
\label{Sbadd}
S_b^{\rm add}(\phi_{\rm D})=\frac{|D_+|^2+|D_-|^2}{D_0^2}
=\frac{|D_+|^2+|D_-|^2}{|D_+ e^{-i\phi_{\rm D}}+D_- e^{i\phi_{\rm
D}}|^2}\,,
\eeq
which usually depends on $\phi_{\rm D}$, unless either
$D_+$ or $D_-$ is zero, which we refer to as the {\it totally
unbalanced} case.  In the case of balanced modulation, when
$|D_+|=|D_-|$, only one quadrature,
\beq
\zeta_{0}^{\rm
balanced}=\frac{\pi}{2}+\frac{1}{2}\left(\arg D_+ + \arg D_-\right)\,,
\eeq
is measured, with additional noise
\beq S_b^{\rm add\,balanced}=\frac{1}{2}\,, \eeq and with a
frequency-independent optimal demodulation phase
\beq
\label{optimalphi}
\phi_{\rm D}^{\rm balanced}=\frac{1}{2}\left[\arg D_{+} -\arg
  D_{-}\right] + N\pi\,,\quad N=0,\,\pm1,\,\ldots.
\eeq
This is the lowest possible additional noise for heterodyne
schemes with just one pair of sidebands. The noise spectral
density can have different shapes as a function of the homodyne
angle~\cite{KLMTV00,BC1}. At different signal sideband
frequencies, the optimal homodyne angle $\zeta_{\rm opt}$ that
gives the lowest homodyne noise can be different.  In homodyne
detection, since both quadratures of the carrier are generally not
available, only a single frequency-independent quadrature can be
measured \footnote{~Unless the output signal is filtered through
the kilometer-scale optical filters proposed by Kimble et al.\
\cite{KLMTV00}.}. By contrast, in heterodyne detection schemes
(except for the balanced case), all quadratures are available for
simultaneous measurement, and the final heterodyne noise at each
frequency will be the minimum of all quadratures.

\section{Noise spectral density and the effect of the additional noise}
\label{sec3}

In this section, we write down  the noise spectral density for both
conventional and RSE interferometers when the RF
modulation-demodulation scheme described in Section~\ref{sec2} is used.

\renewcommand{\arraystretch}{2.5}
\begin{table*}[ht]
\begin{tabular}{ccc}
\hline\hline
Symbol &  Quantity & Expression \\
\hline\hline
$\gamma$ & Half bandwidth of arm cavity &  $\displaystyle \frac{Tc}{4L}$
\\
\hline
$\beta$  & Phase gained by resonant field in arm cavity  & $\arctan(\Omega/\gamma)$  \\
\hline
$h_{\rm SQL}$ &  Free-mass standard quantum limit &
$\displaystyle \sqrt{\frac{8\hbar}{m\Omega^2 L^2}}$  \\
\hline
$I_{\rm SQL}$ &
\renewcommand{\arraystretch}{1}
\begin{tabular}{c} Characteristic input power for conventional \\
  interferometer to reach the SQL at $\Omega=\gamma$ \end{tabular}
\renewcommand{\arraystretch}{2.5}
& $\displaystyle \frac{m L^2\gamma^4}{4\omega_0}$ \\
\hline
$\cal K$ & Coupling constant for radiation-pressure effects  &
$\displaystyle \frac{2(I_0/I_{\rm SQL})\gamma^4}{\Omega^2(\Omega^2+\gamma^2)}$ \\
\hline\hline
\end{tabular}
\caption{Quantities derived from those listed in
Tab.~\ref{tab:basic} \label{tab:def}}
\end{table*}
\renewcommand{\arraystretch}{1}

\subsection{Total noise spectral density}
\label{sec3.1}

The input-output relation for RSE interferometers, including
optomechanical effects, were derived in Refs.~\cite{BC1,BC2}
[see Eqs.~(2.20)--(2.24) and (2.26) of
Ref.~\cite{BC2}]. The output fields in the frequency
band of $(\omega_0-\Lambda,\, \omega_0+\Lambda)$ are (in the conventions used
in this manuscript)
\beq
\left (\matrix{{b}_1^{\omega_0} \cr {b}_2^{\omega_0}}\right)=
\frac{1}{M}\left[e^{2i\beta}\left(\matrix{ C_{11}& C_{12}\cr C_{21} & C_{22}}\right)
\left(\matrix{{a}_1^{\omega_0} \cr {a}_2^{\omega_0} }\right)+
\sqrt{2 {\cal K}}\tau
e^{i\beta}
\left(\matrix{D_1 \cr D_2} \right)\frac{h}{h_{\rm SQL}}\right]\,,
\label{2.20PRD1}
\eeq
where
\beq
M= 1 + \rho^2\, e^{4i\beta}-
2\rho\,e^{2i\beta}\left (\cos{2\phi}+\frac{{\cal K}}{2}\,
\sin{2\phi} \right )\,,
\label{2.21PRD1}
\eeq
\bea
\label{2.22PRD1}
&& C_{11}=C_{22}=(1+\rho^2)\,
\left (\cos{2\phi}+\frac{{\cal K}}{2}\,\sin{2\phi} \right ) -2\rho\,\cos{2\beta}\,,\\
&& C_{12}=-\tau^2\,(\sin{2\phi}+{\cal K}\,\sin^2{\phi})\,, \quad \quad
C_{21}=\tau^2\,(\sin{2\phi}-{\cal K}\,\cos^2{\phi})\,,
\label{2.23PRD1}
\eea
\beq
D_1= - (1+\rho\, e^{2 i\beta})\,\sin{\phi}\,, \quad \quad
D_2= - (-1+\rho\, e^{2 i\beta})\,\cos{\phi}\,.
\label{2.24PRD1}
\eeq
The quantities $\cal K$, $\beta$, $\rho$, $\phi$, $\tau$ and $h_{\rm
SQL}$ are defined in the same way as in Refs.~\cite{BC1,BC2,BC3}.
We denote by $h(\Omega)$ the gravitational strain and
give a summary of the main quantities in Tables~\ref{tab:basic} and
\ref{tab:def}.
We assume that the fields $a^{\omega_0}$ incident on the unused
input of the antisymmetric port are in the vacuum state for all
frequencies. Moreover, the additional heterodyne noise fields
$b^{\omega_0 \pm 2\omega_m}$ in Eq.~(\ref{outQFD}) must also be in
vacuum states, since they are far away from the carrier frequency
and are not affected by the ponderomotive squeezing effects of the
interferometer. We assume that the higher-order terms of the
modulation are not resonant in the interferometer, which is in
general the case. Even if the higher-order sidebands are
resonance, we would not expect any ponderomotive squeezing since
the frequency is too high for the test-mass displacement to
respond to an external force. Using Eqs.~(\ref{33})--(\ref{Sbadd})
and (\ref{2.20PRD1}), we obtain the total heterodyne noise
spectral density in $h$, as a sum of the corresponding homodyne
noise (first term) and the additional heterodyne noise (second
term) [see Eqs.~(\ref{D_0}), (\ref{zeta_0}) for the definitions of
$D_0$ and $\zeta_0$]:
\bea
S_h^{\rm het}(\phi_{\rm D};\Omega)
&=& \frac{h_{\rm SQL}^2}{2{\cal K}}\,
\frac{1}{\tau^2\,\left|D_1\,\sin\zeta_0+D_2\,\cos\zeta_0\right|^2}\,\left [
{\left(C_{11}\,\sin\zeta_0+C_{21}\,\cos\zeta_0\right)^2+
\left(C_{12}\,\sin\zeta_0+C_{22}\,\cos\zeta_0\right)^2} \right ]+  \nonumber \\
&& \frac{h_{\rm SQL}^2}{2{\cal K}}\,
\frac{1}{\tau^2\,\left|D_1\,\sin\zeta_0+D_2\,\cos\zeta_0\right|^2}\,
\left [ \frac{\left|D_+\right|^2+\left|D_-\right|^2} {\left|
D_+\,e^{- i\phi_{\rm D}}+D_-\,e^{i\phi_{\rm
D}}\right|^2}\left|M\right|^2 \right ]\,. \label{s} \eea

We note that the optimal heterodyne noise spectral density at a
given GW signal sideband frequency is the minimum of those
obtained varying $\phi_{\rm D}$ (and thus $\zeta_0$).

\subsection{Conventional interferometers}
\label{sec3.2}
\begin{figure*}
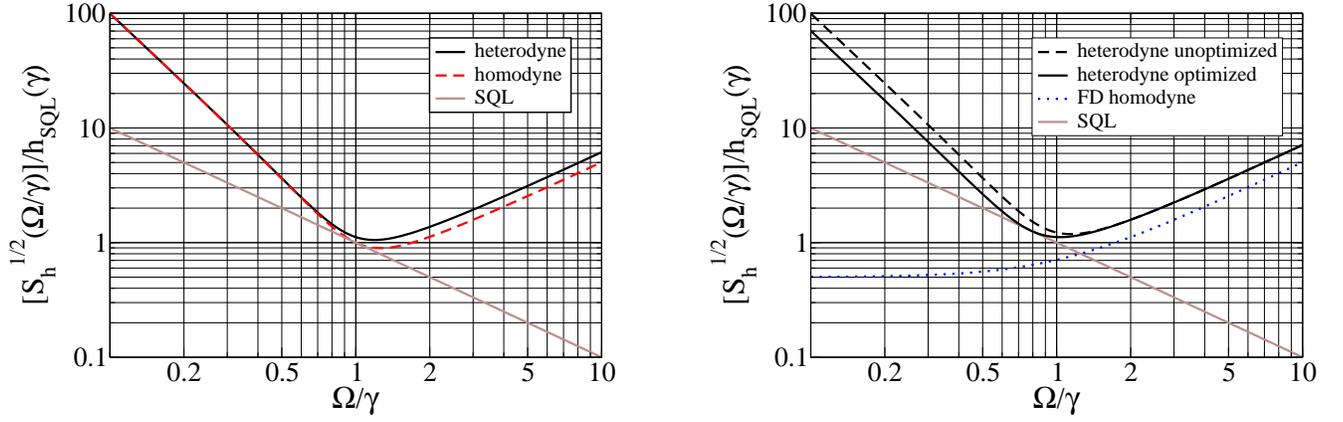

\begin{center}
\begin{tabular}{cc}
\includegraphics[width=0.45\textwidth]{Fig3.eps}
\hspace{0.025\textwidth} & \hspace{0.025\textwidth}
\includegraphics[width=0.45\textwidth]{Fig4.eps}
\end{tabular}
\end{center}
\caption{In the left panel we show the square root of
noise spectral density, in units of
  $h_{\rm SQL}(\gamma)$,  for a conventional
  interferometer with $I_0=I_{\rm SQL}$, using
  balanced heterodyne modulation (solid curve) and the homodyne
  (dashed curve) scheme, plotted as functions of $\Omega/\gamma$.  The second
  quadrature is measured.   In the right panel we plot the noise
  spectral density of the same interferometer, using unbalanced
  heterodyne detection, with homodyne angle chosen at the optimal value for the
  homodyne case, $\zeta(\Omega)=\zeta_{\rm opt\,hom}(\Omega)$ (dashed
  line), and at the re-optimized value for the heterodyne case
  $\zeta(\Omega)=\zeta_{\rm opt\,het}(\Omega)$ (solid curve),
  respectively.  The
  optimal heterodyne noise spectral density {\it without the
  additional noise} is also shown (dotted curve), which agrees with the
  result for frequency-independent homodyne detection~
\cite{KLMTV00}. The SQL line is shown in both panels as gray straight
  lines. \label{fig:rfLIGO-I}}
\end{figure*}
For the power-recycled Fabry-Perot Michelson optical
configuration, the so-called conventional interferometer, the GW
signal appears only in the second (or phase) quadrature.
Furthermore, barring imperfections, the transmission of the
Schnupp sidebands is balanced. In our notation such a scheme is
obtained by setting $D_-=-D_+^*$, with $\phi_{\rm D}=\pm
\pi/2+\arg D_+$ which is the optimal demodulation phase for all frequencies
[see Eq.~(\ref{optimalphi})]. Evaluating Eq.~(\ref{s}) in the case
$\phi=0, \tau=1$, we get
\beq S_h^{\rm het\,\mbox{\scriptsize conv}}= \frac{h_{\rm
SQL}^2}{2{\cal K}} \left({\cal K}^2+1+\frac{1}{2}\right)\,,
\label{sn} \eeq
where the last term inside the parenthesis is the additional
heterodyne noise, which is equal to $1/2$ the shot noise in
homodyne readout scheme (second term), originally derived in
Ref.~\cite{MS91}. In the left panel of Fig.~\ref{fig:rfLIGO-I}, we
plot the noise curves of a conventional interferometer with
$I_0=I_{\rm SQL}$, using homodyne and balanced heterodyne
detection, respectively, with the second quadrature measured. This
is exactly the result in Ref.~\cite{N88}. More sophisticated
modulation schemes that can further lower or eliminate the
additional heterodyne noise {\it in this quadrature} have been
investigated by Schnupp~\cite{Schnupp}, Niebauer et al.\
\cite{N88}, and Meers and Strain~\cite{MS91}.

If, on the contrary, the RF sidebands at the antisymmetric port
are not balanced, one can measure arbitrary quadratures by adjusting
the demodulation phase [see Sec.~\ref{sec2.4}].
As proposed by Vyatchanin, Matsko and
Zubova~ \cite{VMZ}, and further investigated by Kimble, Levin,
Matsko, Thorne and Vyatchanin (KLMTV)~\cite{KLMTV00}, measuring
different quadratures at different GW signal sideband frequencies
can allow conventional interferometers to beat the Standard
Quantum Limit (SQL)~\cite{SQL} significantly, thus converting them
into a QND interferometer.
Somiya~\cite{Somiya} proposed that, by using a frequency-dependent
demodulation phase, a KLMTV-type, frequency-dependent
optimization is achievable in a totally
unbalanced modulation scheme. However, the effect of the
additional heterodyne noise was not explicitly taken into account
and we show in this section that the additional heterodyne noise
plays an important role as soon as one approaches the SQL. So much
so, that for totally unbalanced heterodyne detection, the SQL
cannot be beaten, and for intermediate levels of imbalance the SQL
is beaten by very modest amounts.

For simplicity, we first consider a totally unbalanced modulation
scheme (which was the case investigated by Somiya~\cite{Somiya}), in which only
$D_+$ (or only $D_-$) is non-zero. From Eq.~(\ref{s}), fixing
$\tau=1$, $\rho=0$ and $\phi=0$, we have
\beq
\label{Shetconv}
S_h^{\rm
het\,conv}=\frac{h_{\rm SQL}^2}{2{\cal K}}\left[({\cal
K}-\tan\zeta_0)^2+1+\frac{1}{\cos^2\zeta_0}\right]\,,
\eeq
where the
last term inside the parenthesis is the additional noise due to
heterodyne detection. Using the optimal detection angle in the
(frequency dependent)
homodyne case~\cite{VMZ,KLMTV00}, $\zeta_{\rm opt\,hom}=\arctan{\cal K}$, one has
\beq
\label{shhethom}
S_h=\frac{h_{\rm SQL}^2}{2{\cal K}}\left({\cal K}^2+2\right) \ge \sqrt{2} h_{\rm SQL}^2\,,
\eeq
which cannot reach the SQL. Re-optimizing the detection angle, we
obtain $\zeta_{\rm opt\,het}=\arctan({\cal K}/2)$. This gives
\beq
\label{shhethet}
\left[S_h^{\rm het\,conv}\right]_{\rm opt}=\frac{h_{\rm SQL}^2}{2{\cal
    K}}\left(\frac{{\cal K}^2}{2}+2\right)\ge h_{\rm SQL}^2\,, \eeq
    which only touches, but never beats, the SQL. In the right panel
    of Fig.~\ref{fig:rfLIGO-I}, we plot the noise curve of a
    conventional interferometer with $I_0=I_{\rm SQL}$, the heterodyne
    noise spectral density using $\zeta_{\rm opt\,hom}$
    [given by Eq.~(\ref{shhethom})], and the optimal heterodyne noise spectral
    density [given by Eq.~(\ref{shhethet})]. As can be further verified, having
    two sidebands with unequal amplitude can allow the interferometer
    to beat the SQL, but only by very moderate amounts, and in limited
    frequency bands.

We might still expect to use more sophisticated
modulation-demodulation schemes to lower the additional heterodyne
noise while retaining the possibility of variable-quadrature
optimization.  However, as we shall see in Sec.~\ref{sec4}, such
an effort will be significantly limited by the Heisenberg
uncertainty principle.
\begin{figure*}
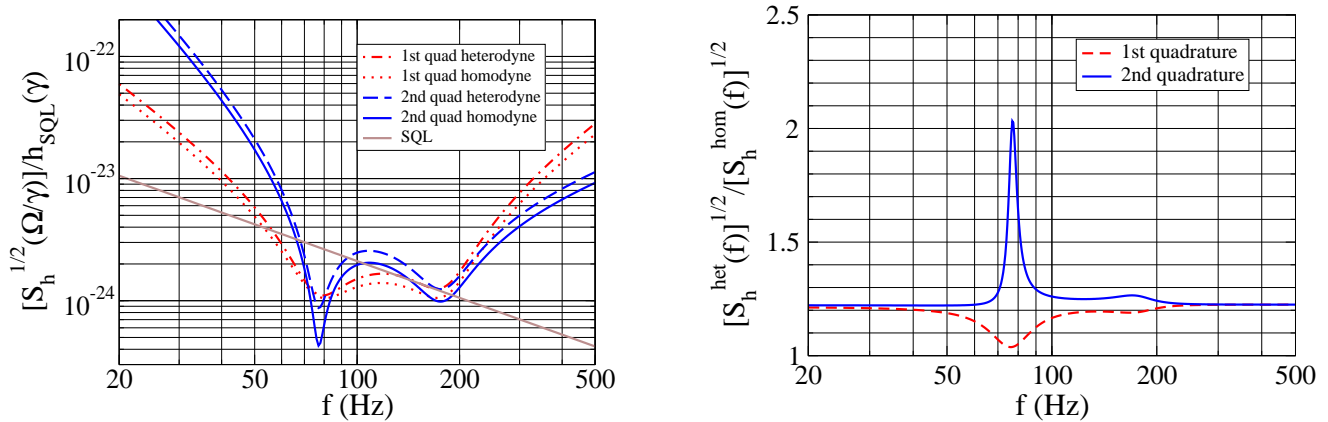

\begin{center}
\begin{tabular}{cc}
\includegraphics[height=0.3\textwidth,width=0.45\textwidth]{Fig5.eps}
\hspace{0.025\textwidth} & \hspace{0.025\textwidth}
\includegraphics[height=0.31\textwidth,width=0.45\textwidth]{Fig6.eps}
\end{tabular}
\end{center}
\caption{A detuned RSE interferometer ($T=0.033$, $\rho=0.9$, $\phi=\pi/2-0.47$,
  $m=30\,{\rm kg}$, $I_0=I_{\rm SQL}$) using balanced heterodyne detection.
  In the left panel we plot the square root of total
  heterodyne noise spectral density in the first (dash-dot curve) and
  second (long dashed curve) quadratures, compared with the homodyne ones
  (dot curve and solid curve, respectively).
  The SQL line is also shown as a gray straight line.
  In the right panel
  we show the ratio of the square roots of the heterodyne and the homodyne
  noise spectral densities,
  for the first (dashed curve) and second (solid curve)
  quadratures.
\label{fig:rfSRbal}}
\end{figure*}

\subsection{Signal-recycled interferometers}
\label{sec3.3}

In this section, we give some examples of noise curves of detuned RSE
interferometers with a heterodyne readout scheme, and compare them
to the homodyne cases.

In the balanced scheme the additional heterodyne noise is the
lowest, but only one quadrature can be measured. For this case we
show the effect of the additional heterodyne noise on the
sensitivity curves in Fig.~\ref{fig:rfSRbal}. In the left panel,
we plot  the noise curves for a detuned RSE interferometer with $T=0.033$,
$\rho=0.9$, $\phi=\pi/2-0.47$, $I_0=I_{\rm SQL}$ and $m=30\,{\rm
kg}$, (the configuration considered in Refs.~\cite{BC1,BC2,BC3})
when the first ($\zeta=\pi/2$) and second quadratures ($\zeta=0$)
are measured, by homodyne and balanced heterodyne read-out
schemes. In the right panel, we plot the ratio of the heterodyne
noise to the corresponding homodyne ones. The additional
heterodyne noise has more features around the two valleys of the
noise curves, where the optomechanical dynamics (the RSE transfer
function) determines the shape of the curves. Above $\sim 200$ Hz,
the ratio between the square roots of the heterodyne and the
homodyne noise spectral densities assumes the constant value
$\sqrt{3/2} \sim 1.22$, which is due to the additional heterodyne
noise when the shot noise dominates [see Eq.~(\ref{sn})].

\begin{figure*}
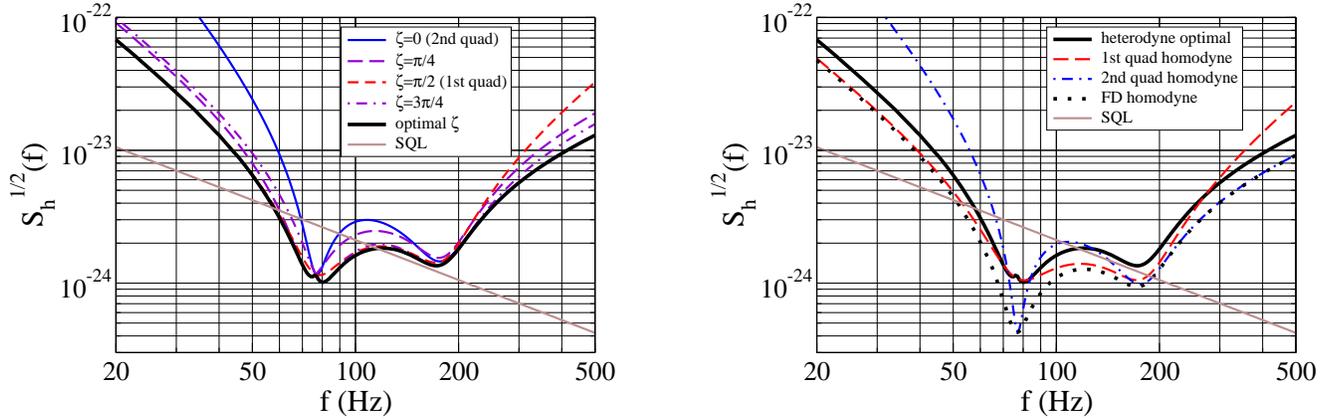

\begin{center}
\begin{tabular}{cc}
\includegraphics[width=0.45\textwidth]{Fig7.eps}
\hspace{0.025\textwidth} & \hspace{0.025\textwidth}
\includegraphics[width=0.45\textwidth]{Fig8.eps}
\end{tabular}
\end{center}
\caption{A detuned RSE interferometer ($T=0.033$, $\rho=0.9$, $\phi=\pi/2-0.47$,
  $m=30\,{\rm kg}$, $I_0=I_{\rm SQL}$, same as Fig.~\ref{fig:rfSRbal})
  using totally unbalanced heterodyne detection. In the left panel we show
  the noise curves for quadratures with $\zeta=0$ (solid curve), $\pi/4$ (long
  dashed curve),
  $\pi/2$ (short dashed curve) and $3\pi/4$ (dash-dot curve)
  are shown, together with the final heterodyne noise optimized at all
  sideband frequencies (thick solid curve). In the right panel, the final
  heterodyne noise (thick solid curve) is shown along with the homodyne noise in the
  first (dashed curve) and second (dash-dot curve) quadratures.
  The optimal
  heterodyne noise without the additional noise is also shown (dotted
  curve) for comparison. The SQL is plotted in both panels as gray
  straight lines.  \label{fig:rfSRun}}
\end{figure*}

Practical implementation of the RF sidebands in the interferometer
has shown that detuned RSE configurations are likely to be very
unbalanced~\cite{MW,BCM2}. In the left panel of
Fig.~\ref{fig:rfSRun}, we plot the unbalanced heterodyne noise
spectral densities for the same interferometer parameters used in
Fig.~\ref{fig:rfSRbal}, with $\zeta_0=0$, $\pi/4$, $\pi/2$ and
$3\pi/4$, and the optimal heterodyne noise obtained by maximizing
over $\zeta_0$ at each sideband frequency. Indeed, in the
heterodyne readout scheme we have the advantage of optimizing the
detection angle at different frequencies. At each particular
signal sideband frequency, the optimal heterodyne noise spectral
density is just the minimum of all quadratures. In the right panel
of Fig.~\ref{fig:rfSRun}, we compare the optimal heterodyne noise
with the homodyne noise at $\zeta=0$ and $\zeta=\pi/2$. As we see
from this example, for the same interferometer configuration,
neither the homodyne nor the heterodyne readout can provide a
noise spectral density that is the lowest for all GW signal
sideband frequencies. To make a more rigorous comparison between
these two schemes a more critical study is required that takes
into consideration specific astrophysical GW sources, the
experimental feasibility and the other sources of noise, as well.
[As an example, current Advanced LIGO design estimates  the
dominant, thermoelastic component at about the SQL~\cite{therm}.
To lower the thermoelastic contribution below the SQL an
interesting and challenging proposal has been analyzed recently~\cite{DOSTVB}.]
The optimization of homodyne versus heterodyne
readout schemes which include those effects is currently underway,
and will be reported elsewhere~\cite{BCM2}.

As shown in Ref.~\cite{BC3}, detuned RSE interferometers have an
unstable optomechanical resonance. In the parameter regime
emphasized in Refs.~\cite{BC1,BC2,BC3,BC4}, the unstable resonance
lies within the observation band --- which gives a dip in the
noise spectrum. Consequently, the control scheme must sense and
act on the motion of the system within the observation band. In
Ref.~\cite{BC3}, an idealized control scheme is conceived for the
homodyne readout, which suppresses the instability and leaves the
noise spectral density unchanged. The same control issue will need
to be addressed with the heterodyne readout scheme as well.

\section{More general discussion of heterodyne schemes: minimal additional noise
and quantum limit}
\label{sec4}

In Sec.~\ref{sec2} we discussed the sinusoidal modulation-demodulation scheme,
which is the easiest to implement. There exist more
sophisticated schemes, such as those proposed by Schnupp, investigated by
Niebauer et al.\ \cite{N88}, and Meers and Strain~\cite{MS91},
that can further optimize the interferometer
performances. These authors restricted their analyses to low-power interferometers
and focused on the detection of the second (or phase) quadrature.  In this section, we
extend their discussions to the more general case where all quadratures
can be measured. As we shall see, although modulation/demodulation readout schemes
offers the advantage of variable-quadrature optimization, they
are in general limited in converting non-QND interferometers to
(broadband) QND interferometers.

\subsection{Quantum Limit for the additional heterodyne noise}
\label{sec4.1}

The field coming out from the dark port can be written, in time domain,
as
\beq
\label{E}
E(t)=\left[A(t)\cos\omega_0 t+P(t)\sin\omega_0 t\right] +
\left[E_1(t)\cos\omega_0 t + E_2(t)\sin\omega_0 t\right]\,,
\eeq
where the first term is the transmitted Schnupp sideband fields in
the form of a combination of amplitude modulation [$A(t)$] and phase
modulation [$P(t)$] to the carrier. In Eq.~(\ref{E})
we denoted by $E_1(t)$ and $E_2(t)$
the quadrature fields containing GW signal and quantum fluctuations.
The output from the photodetector is then
\beq
\label{PDoutquadrature}
i(t)\propto A(t) E_1(t)+ P(t) E_2(t)\,.
\eeq
The amplitude and phase modulation is in general periodic
functions, with the same angular frequency $\omega_{\rm m}$
[$\omega_{0}¥ \gg \omega_{\rm m} \gg \Omega_{\rm GW}$]:
\bea
A(t)&=&\sum_{k}A_{k}e^{-i k \omega_{m} t}\,,\quad A_{k}=A^{*}_{-k}¥\,; \\
P(t)&=&\sum_{k}P_{k}e^{-i k \omega_{m} t}\,,\quad P_{k}=P^{*}_{-k}\,.\\
\eea
In the frequency domain Eq.~(\ref{PDoutquadrature}) reads:
\beq
i(\Omega) \propto
\sum_{k}
\left[
A_{k} b_{1}^{\omega_{0}}(\Omega-k\omega_{\rm m})
+
P_{k} b_{2}^{\omega_{0}}(\Omega-k\omega_{\rm m}¥)
\right]\,.
\eeq
Denoting the demodulation function with $D(t)$, the demodulated output is:
\beq
O(t)=D(t)i(t)\propto D(t)A(t) E_{1}(t)+D(t)P(t) E_{2}(t)\,.
\label{DO}
\eeq
The demodulation function $D(t)$ should have the same frequency as
the modulation functions, therefore:
\beq
D(t)=\sum_{k}D_{k}e^{-i k \omega_{m} t}\,,\quad D_{k}=D^{*}_{-k}¥\,.
\eeq
Note that Eq.~(\ref{DO}) is a generalization of Eq.~(4) of
Ref.~\cite{MS91}. Using the above equations, the Fourier transform
of the demodulated output (\ref{DO}) can be written as:
\beq
\label{generalout}
\widetilde{O}(\Omega)
=\sum_{k} D_{k}^{*}
\left[A_{k} b_{1}^{\omega_{0}}(\Omega)
+P_{k} b_{2}^{\omega_{0}¥}¥(\Omega)\right]
+
\sum_{p \neq 0} \sum_{k}
D_{k}^{*}
\left[
A_{k+p}b_{1}^{\omega_{0}¥}¥(\Omega-p\,\omega_{m}¥)
+
P_{k+p}b_{2}^{\omega_{0}¥}¥(\Omega-p\,\omega_{m}¥)
\right]\,.
\eeq
Let us suppose that the low-frequency component of
$\widetilde{O}(\Omega)$ is filtered out, then the first term in
Eq.~(\ref{generalout}) gives a frequency-independent quadrature
field near $\omega_{0}$, while the second term gives the
additional heterodyne noise that arises from quantum fluctuations
near $\omega_{0}¥ \pm p \, \omega_{\rm m}$, with $p=\pm 1$, $\pm
2$, $\pm 3$, \ldots. Since $\omega_{\rm m} \gg \Omega_{\rm GW}$,
these fields are not affected by ponderomotive squeezing effects
in the interferometer arm cavities and will be in the vacuum
state. As a consequence, the additional heterodyne noise will also
be frequency independent. [Unless frequency-dependent squeezed
states are injected into the dark port of the interferometer.] In
this way, for any particular quadrature $\zeta$, there is a
uniform minimum of the additional heterodyne noise at all
frequencies.

Let us now construct for an arbitrary
quadrature $\zeta$ the optimal demodulation function $D(t)$ and evaluate the minimal
additional noise. If we want to measure $b^{\omega_{0}}_{\zeta}$,
Eq.~(\ref{generalout}) says that we have to impose
\beq
\label{Dfreq}
\left(\sum_{k}¥D_{k}^{*}A_{k},\sum_{k}¥D_{k}^{*}P_{k}\right)
=
(\sin\zeta,\cos\zeta)\,,
\eeq
or in the time domain,
\beq
\label{constrainD}
(\sin\zeta,\cos\zeta)=\left(
\frac{1}{T}\int_0^{T} D(t)A(t) dt,
\frac{1}{T}\int_0^{T} D(t)P(t) dt
\right)\,,
\eeq
where $T=2\pi/\omega_{\rm m}$ is the common period of the
modulation and demodulation functions. Note that, in order for the
quadrature $\zeta$ to be measured, Eqs.~(\ref{Dfreq}) and
(\ref{constrainD}) need only be true up to a constant factor.
Having written them in the current way, we have in fact chosen a
specific normalization for $D(t)$. Using the Parseval theorem and
Eq.~(\ref{constrainD}), we derive for the spectral density of the
additional heterodyne noise:
\bea
S^{\rm add}&=&
\sum_{p\neq 0}
\left|\sum_{k}D^{*}_{k}¥A_{k+p}\right|^{2}
+
\sum_{p\neq 0}
\left|\sum_{k}D^{*}_{k}P_{k+p}\right|^{2}\,,
\nonumber \\
&=&
\sum_{p}
\left|\sum_{k}D^{*}_{k}¥A_{k+p}\right|^{2}
+
\sum_{p}
\left|\sum_{k}D^{*}_{k}P_{k+p}\right|^{2}¥
-
\left|\sum_{k}D^{*}_{k}A_{k}\right|^{2}
-
\left|\sum_{k}D^{*}_{k}P_{k}\right|^{2}¥\,, \nonumber \\
&=&
\frac{1}{T}\int_{0}^{T} D^{2}(t)\left[A^{2}(t)+P^{2}¥(t)\right]dt
-\left(\frac{1}{T}\int_{0}^{T} D(t) A(t) dt\right)^{2}
-\left(\frac{1}{T}\int_{0}^{T} D(t) P(t) dt\right)^{2}¥\,, \nonumber \\
&=& \frac{1}{T}\int_{0}^{T} D^{2}(t)\left[A^{2}(t)+P^{2}¥(t)\right] dt
-1 \,.
\label{generalSadd}
\eea
[Note that Eq.~(\ref{generalSadd}), which corresponds directly to Eq.~(12) of
Ref.~\cite{MS91}, is also consistent with Eq.~(18) of
Ref.~\cite{MS91}, since Eqs.~(\ref{Dfreq}) and (\ref{constrainD}) have
already imposed a normalization for $D(t)$.]  In order to find the
$D(t)$ that satisfies Eq.~(\ref{constrainD}) and minimize
$S^{\rm add}¥$, we introduce two Lagrange multipliers, $\lambda$ and
$\mu$, and impose,

\beq
\label{variation}
\delta\int dt
\Bigg\{\left[(A^{2}(t)+P^{2}(t)\right]D^{2}(t)- 2 \lambda A(t) D(t) -
2 \mu
P(t) D(t) \Bigg\} =0\,,
\eeq

which yields:
\beq
\label{optimalD}
D(t)=\frac{\lambda A(t) +\mu P(t) }{A^{2}(t)+P^{2}(t)}\,.
\eeq
[In Eq.~(\ref{variation}), the factors of 2 in front of $\lambda$ and $\mu$ are added for simplicity.]
Inserting Eq.~(\ref{optimalD}) back into Eq.~(\ref{constrainD}) gives,
\beq
\label{lambdamu}
{\bf M}
\left(
\begin{array}{c}
    \lambda \\
    \mu
\end{array}
\right)
=
\left(
\begin{array}{c}
    \sin\zeta \\
    \cos\zeta
\end{array}
\right)\,,
\eeq
where
\beq
{\bf M} =\left(
\begin{array}{cc}
    \displaystyle \frac{1}{T}\int_{0}^{T} \frac{A^{2}(t)}{A^{2}(t)+P^{2}¥(t)} dt
    &
    \displaystyle \frac{1}{T}\int_{0}^{T} \frac{A(t)P(t)}{A^{2}(t)+P^{2}¥(t)} dt
\\
    \displaystyle \frac{1}{T}\int_{0}^{T} \frac{A(t)P(t)}{A^{2}(t)+P^{2}¥(t)} dt
    &
    \displaystyle \frac{1}{T}\int_{0}^{T} \frac{P^{2}(t)}{A^{2}(t)+P^{2}¥(t)} dt
\end{array}
\right)\,.
\eeq
The optimal demodulation function for the $\zeta$ quadrature is then
given by inverting Eq.~(\ref{lambdamu}) and inserting the resulting
$\lambda$ and $\mu$ into Eq.~(\ref{optimalD}). The minimal additional
noise can be then obtained by inserting the optimal demodulation function
into Eq.~(\ref{generalSadd}):

\beq
S^{\rm add\,min}(\zeta)=
\left(\begin{array}{cc}
    \lambda & \mu
\end{array}\right)
{\bf M}
\left(
\begin{array}{c}
    \lambda \\
    \mu
\end{array}
\right)
-1
=\left(\begin{array}{cc}
    \sin\zeta & \cos\zeta
\end{array}\right)
{\bf M}^{-1}
\left(
\begin{array}{c}
    \sin\zeta \\
    \cos\zeta
\end{array}
\right)
-1
=\left(\begin{array}{cc}
    \sin\zeta & \cos\zeta
\end{array}\right)
\left({\bf M}^{-1}¥-{\bf I}\right)
\left(
\begin{array}{c}
    \sin\zeta \\
    \cos\zeta
\end{array}
\right)\,.
\eeq

Moreover, we note an interesting property of ${\bf M}$:
\beq
{\bf I}-{\bf M}=(\det{\bf M}){\bf M}^{-1}\,.
\eeq
As a consequence,
\beq
{\bf M}^{-1}-{\bf I}={\bf M}^{-1}({\bf I}-{\bf M})=(\det{\bf M})({\bf M}^{-1})^{2},
\eeq
so
\beq
\det\left[{\bf M}^{-1}-{\bf I}\right]=1\,.
\eeq
This implies that, the minimal additional noise can be written in a
form
\beq \label{Saddsqz} S^{\rm add\,min}(\zeta) = \left(
\begin{array}{cc}
\sin\zeta & \cos\zeta
\end{array}
\right)
\left(
\begin{array}{rr}
\cos\phi & \sin\phi \\
-\sin\phi & \cos\phi
\end{array}
\right)
\left(
\begin{array}{cc}
e^R &  \\
 & e^{-R}
\end{array}
\right)
\left(
\begin{array}{rr}
\cos\phi & -\sin\phi \\
\sin\phi & \cos\phi
\end{array}
\right)
\left(
\begin{array}{c}
\sin\zeta \\
\cos\zeta
\end{array}
\right)\,, \eeq
with $\phi$ and $R$ frequency-independent, and determined by the
eigenvectors and eigenvalues of the matrix ${\bf M}^{-1}-{\bf 1}$,
which are determined ultimately by the amplitude and phase
modulations. It is interesting to note that this minimal noise
spectrum is of exactly the same form as that of a squeezed state.

This phenomenon could in fact be anticipated from quantum mechanics. For
the same sideband frequency $\Omega$, the different quadratures do not
commute with each other, and have the following commutation relations:
\beq
\label{quadraturecomm}
\left[b^{\omega_0}_{\zeta}(\Omega),b^{\omega_0\,\dagger}_{\zeta'}(\Omega')\right]=
    2\pi i \sin(\zeta-\zeta')\delta(\Omega-\Omega')\,.
\eeq
As a consequence, quantum fluctuations in the various quadratures are
constrained by the Heisenberg uncertainty principle. As is well known,
the squeezed states have the minimum noise spectrum allowed by the
uncertainty principle. In modulation/demodulation schemes, all
quadratures can be read out, with additional noise:
\beq
b^{\rm het}(\zeta,\Omega) = b^{\omega_{0}}_{\zeta}(\Omega) +
n(\zeta,\Omega)\,.
\eeq
So, all output observables should commute with each other, and as
a consequence
\beq
\left[b^{\rm het}(\zeta,\Omega),b^{\rm het \dagger}(\zeta',\Omega')\right]=0\,.
\eeq
Since $b^{\omega_{0}}_{\zeta}(\Omega)$ and $n(\zeta,\Omega)$ come from
different frequency bands of the output field, they must commute with
each other, so we must have that the mutual commutators of $n(\zeta,\Omega)$ cancel
those of $b^{\omega_{0}}_{\zeta}(\Omega)$:
\beq
\label{noisecomm}
\left[n(\zeta,\Omega),n^{\dagger}¥(\zeta',\Omega')\right]=
- 2\pi i \sin(\zeta-\zeta')\delta(\Omega-\Omega')\,.
\eeq
Since they do not commute with each other, the additional noise $n(\zeta,\Omega)$
is also subject to the constraint of the Heisenberg uncertainty principle ---
in the same way as $b^{\omega_{0}}_{\zeta}(\Omega)$,
since the commutators only differ by a sign [see
Eqs.~(\ref{quadraturecomm}) and
(\ref{noisecomm})]. This explains why the minimum additional heterodyne noise has a spectral
density of the same form as the squeezed states.
The minimum noise spectrum (\ref{Saddsqz}) can be regarded as a {\it quantum
limit} for modulation/demodulation schemes.

\subsection{Impact of the Quantum Limit on conventional interferometers}
\label{sec4.2}

As discussed in Refs.~\cite{VMZ,KLMTV00}, using an appropriate
readout scheme, conventional interferometers can achieve QND
performance through a cancellation between shot and
radiation-pressure noises. If the quadrature $\zeta$ is
measured, we have
\beq
\label{cancellation}
b^{\rm
het\,conv}(\zeta,\Omega)=\cos\zeta \left[e^{2 i
    \beta}\left(\tan\zeta-\mathcal{K}\right)a_1^{\omega_0}+e^{2 i
    \beta}a_2^{\omega_0}+e^{i \beta}\sqrt{2\mathcal{K}}\frac{h}{h_{\rm
    SQL}}\right]+n(\zeta,\Omega)\,,
\eeq
and if we choose to measure the quadrature with $\zeta=\zeta_{\rm c}=\arctan\mathcal{K}$
\footnote{Note that $\zeta = \zeta_c$ is not the optimal quadrature.} the part of the shot noise
[the term proportional to $a_1^{\omega_0} \tan\zeta$ inside the bracket of
Eq.~(\ref{cancellation})] cancels the radiation-pressure noise
[the term proportional to $\mathcal{K} a_1^{\omega_0}$ inside the
bracket of Eq.~(\ref{cancellation})].
The remaining shot noise [obtained from
the term proportional to $a_2^{\omega_0}$ inside the
bracket of Eq.~(\ref{cancellation})], normalized to
unit signal strength, is inversely proportional to
$\mathcal{K}$, and it can be made lower (eventually lower than the SQL noise)
by taking larger $\mathcal{K}$. However, for larger values of $\mathcal{K}$,
$\tan\zeta_{\rm c}$ grows and the corresponding $\cos\zeta_{\rm c}$ decreases.
As can be seen from Eq.~(\ref{cancellation}), this implies
an even smaller signal strength in the detected quadrature, which
makes the additional noise, $n(\zeta_{\rm c},\Omega)$, more and more important.
In fact, more generally the additional noise limits the extent to which the
interferometer can beat the SQL. Writing the total heterodyne noise
spectral density [of which Eq.~(\ref{Shetconv}) is a special case], as
\beq
S_h=\frac{h_{\rm
SQL}^2}{2\,\mathcal{K}}\left[(\mathcal{K}-\tan\zeta)^2+1+\frac{S^{\rm
add}(\zeta)}{\cos^2\zeta}\right]\,,
\label{nsd}
\eeq
and following the argument that had led us to Eq.~(\ref{shhethet}),
we obtain the following lower limit for the heterodyne noise:
\beq
\label{opt}
\left[S_h(\Omega)\right]_{\rm opt}\ge \sqrt{
S^{\rm
add\,min}(\zeta_{\rm opt}(\Omega))}\,
h_{\rm SQL}^2\,,
\eeq
where $\zeta_{\rm opt}(\Omega)$ is the optimal detection quadrature
at frequency $\Omega$, which depends also on the shape of
$S^{\rm add}(\zeta)$.
Equation (\ref{opt}) says that, in order
to beat the SQL significantly, the additional heterodyne noise at the
optimal quadrature has to be much smaller than unity.  However, since
the additional heterodyne noise is frequency independent, and subject
to the quantum limit (\ref{Saddsqz}), this requirement cannot always
be fulfilled if the optimal homodyne quadrature varies significantly
with frequency in the observation band. As a consequence, heterodyne
schemes will have very limited power in converting conventional
interferometers into (broadband) QND interferometers.

Due to the simplicity of the input-output relations of
conventional interferometers, we can go a step further and obtain
a cleaner result in this case. Let us suppose that the additional
heterodyne noise have exactly the form of Eq.~(\ref{Saddsqz}),
with generic values of $\phi$ and $e^{R}$, i.e.\ it is {\it
quantum limited}. Inserting Eq.~(\ref{Saddsqz}) into
Eq.~(\ref{nsd}), we find the frequency-dependent optimal detection
phase,
\beq
\label{optimalconv}
\tan\zeta_{\rm opt}(\Omega)=\frac{[1-\tanh^2(R/2)]\,\mathcal{K}(\Omega)+
2\tanh(R/2)\sin2\phi}{2+2\tanh(R/2)\cos2\phi}\,,
\eeq
and obtain:
\beq
\label{Shgenericconv}
S_h^{\rm quant\, lim}(\Omega)=
\left[
\frac{[1+2\cos(2\phi)\tanh(R/2)+\tanh^2(R/2)]\,\mathcal{K}^2(\Omega)-
4\sin(2\phi)\tanh(R/2)\,\mathcal{K}(\Omega)+4}
{4\mathcal{K}(\Omega)\,[1+\cos(2\phi)\tanh(R/2)]}
\right]
h_{\rm SQL}^2(\Omega)\,.
\eeq
Moreover, the quantum-limited heterodyne noise spectral density
(\ref{Shgenericconv}) can be recast into exactly the same form as
that of a frequency-independent homodyne detection
\beq
\label{Shhetequiv}
S_h^{\rm quant\, lim}(\Omega)=\frac{h_{\rm SQL}^2(\Omega)}{2\mathcal{K}_{\rm eff}(\Omega)}\,
\left[(\mathcal{K}_{\rm eff}(\Omega)-\tan\zeta_{\rm eff})^2+1\right]\,,
\eeq
with
\beq
\label{Keffzeff}
\tan\zeta_{\rm eff} \equiv \frac{\tanh(R/2)\sin(2\phi)}{1+\tanh(R/2)\cos(2\phi)}\,,\quad
\mathcal{K}_{\rm eff}(\Omega) \equiv
\left[1-\frac{1-\tanh^2(R/2)}{2+2\tanh(R/2)\cos(2\phi)}\right]\,\mathcal{K}(\Omega)\,.
\eeq
Note that in the definition of $\mathcal{K}_{\rm eff}$ the
quantity multiplying $\mathcal{K}$ (which is less than 1, since
$-1<\tanh(R/2)<+1$) can be absorbed into the input power [see the
definition of $\mathcal{K}$ in Table~\ref{tab:def}]. Equations
(\ref{Shhetequiv}) and (\ref{Keffzeff}) therefore relate a
conventional interferometer with a quantum limited heterodyne
readout scheme to an identical conventional interferometer, but
with lower input power and a frequency-independent homodyne
readout scheme. As discussed by KLMTV, the latter does not exhibit
broadband QND behaviour [although fine tunings of parameters can
sometimes give a moderate SQL-beating noise spectral density].
This means, that the variable-quadrature optimization provided by
heterodyne readout schemes does not enhance the QND performance of
conventional interferometers at all.

Nevertheless, as the equivalence also suggests, quantum-limited
heterodyne detection does not deteriorate the sensitivity with
respect to frequency-independent homodyne detection, except for
the lower effective optical power, which can in principle be made
as close as possible to the true optical power, as $e^R\rightarrow
+\infty$. For certain specially designed interferometers, such as
the speed-meter interferometers \cite{Speedmeters} with
Michelson~\cite{PC02} or Sagnac~\cite{C02,K02} topologies, the
optimal homodyne angle is largely constant over a broad frequency
band.  These interferometers already exhibit broadband QND
behavior with frequency-independent homodyne detection. In this
situation, a heterodyne detection scheme (e.g., the Schnupp
square-wave demodulation scheme), optimized for that particular
quadrature, can be employed, e.g., for technical reasons, without
compromising the sensitivity.

\section{Conclusions}
\label{sec5}

In this paper, we applied a quantum optical formalism to a
heterodyne readout scheme for advanced GW interferometers such as
Advanced LIGO. Our results provide a foundation for the
astrophysical optimization of Advanced LIGO interferometers and
should be used to decide whether a homodyne or heterodyne readout
scheme is more advantageous.

One of the advantages of the heterodyne readout scheme (with the
exception of balanced heterodyning), is that all output
quadratures are available for measurement, providing a way of
optimizing the sensitivity at each frequency. This result cannot
be easily achieved in homodyne detection. However, as originally
discovered by Gea-Banacloche and Leuchs~\cite{GBL} and by Schnupp~\cite{Schnupp},
and analyzed by Niebauer et al.\ \cite{N88}, and Meers and Strain~\cite{MS91}
in the low-power limit, heterodyne detection leads to an additional noise term
which is a direct and necessary consequence of the Heisenberg
uncertainty principle.

In the specific case of detuned RSE interferometers planned for
Advanced LIGO, we derived the expressions for the total heterodyne
noise spectral density [see Eqs.~(\ref{2.21PRD1})--(\ref{s}),
(\ref{zeta_0})], assuming a pair of Schnupp sidebands with
arbitrary amplitude ratios. In the balanced case the effect of the
additional heterodyne noise is shown in Fig.~\ref{fig:rfSRbal}. In
the more practical very unbalanced~\cite{BCM2} configuration, we
compared the noise curve in the optimal heterodyne case, obtained
by maximizing over the heterodyne phase at each sideband
frequency, with some noise curves obtained when the homodyne
readout scheme is used. The results are shown in
Fig.~\ref{fig:rfSRun}. Neither the homodyne nor the heterodyne
readout provides a noise spectral density that is the lowest for
all frequencies. Moreover, the differences between the noise
curves occur mainly in the frequency band $70\mbox{--}200$ Hz
where other sources of noise in Advanced LIGO will probably
dominate, e.g.,  thermal noise~\cite{therm} (unless more
sophisticated techniques are implemented~\cite{DOSTVB}). So,
before drawing any conclusion on which readout scheme is
preferable, the comparison between them must take into account the
other sources of noise present in Advanced LIGO and should be
addressed with reference to specific astrophysical GW sources,
such as neutron-star and/or (stellar mass) black-hole binaries,
for which the GW spectrum is a power law with an upper cutoff ranging
from $\sim 200$\,Hz to several kHz,
and also low-mass X-ray binaries which
require narrowband configurations (detuned RSE) around $500-700$
Hz. In this paper we have provided a framework in which these
optimizations can be carried out. We shall report on the results
of the optimization elsewhere~\cite{BCM2}.

From a more theoretical point of view, we worked out a
frequency-independent quantum limit for the additional heterodyne
noise [see Eq.~(\ref{Saddsqz})], which made more explicit the
following fact: lowering the additional heterodyne noise while
simultaneously retaining the ability to measure more than one
quadrature is incompatible in heterodyne detection, which is
inherently frequency-independent unless frequency-dependent
squeezing techniques are implemented. In particular, this
incompatibility seriously limits the extent to which conventional
interferometers can beat the SQL using heterodyne readout scheme.
Indeed, we show in Sec.~\ref{sec4.2} that conventional interferometers with
quantum limited heterodyne detection are equivalent to conventional
interferometers with frequency-independent homodyne detection and lower optical
power. However, for third-generation GW
interferometers with speedmeter-type configurations~\cite{Speedmeters,PC02,C02,K02},
which are already QND interferometers under an appropriate frequency-independent
homodyne detection, heterodyne readout schemes can in principle be employed
without compromising their sensitivity.

\acknowledgments
We thank Peter Fritschel, James Mason and Ken Strain for stimulating discussions, and
Kip Thorne for his continuous encouragement and for very useful interactions.
We also thank Peter Fritschel and Ken Strain for drawing our attention to
the advantages of variable quadrature detection in heterodyne schemes.

We acknowledge support from National Science Foundation grant
PHY-0099568 (AB and YC) and PHY-0107417 (NM). The research for AB
was also supported by Caltech's Richard Chace Tolman Fund. The
research for YC was also supported by the David and Barbara Groce
Fund at the San Diego Foundation. Part of YC's contribution to
this work was made while visiting the GW research group at the
Australian National University. The author is grateful for their
support.

\end{document}